\RequirePackage{fix-cm}
\documentclass[pdftex,twocolumn]{svjour3}
\smartqed  
\usepackage{graphicx}
\usepackage{amsmath}
\usepackage{amsfonts}
\usepackage{natbib}

\spnewtheorem{thm}{Theorem}{\bf}{\rm}
\spnewtheorem*{algorithm}{Algorithm}{\bf}{\rm}
\spnewtheorem*{thmn}{Theorem}{\bf}{\rm}

\begin{document}

\title{Maximum likelihood estimation of the Fisher--Bingham distribution via efficient calculation of its normalizing constant
}

\titlerunning{Maximum likelihood estimation of the Fisher--Bingham distribution}        

\author{Yici Chen         \and
         Ken'ichiro Tanaka 
}


\institute{Y. Chen \at
              Department of Information Science and Technology, The University of Tokyo \\
              \email{cyc261107@gmail.com}           
           \and
           K. Tanaka \at
              Department of Mathematical Informatics, The University of Tokyo, Tokyo, Japan \\
              \email{kenichiro@mist.i.u-tokyo.ac.jp}  
}

\date{Received: date / Accepted: date}

\maketitle

\begin{abstract}
This paper proposes an efficient numerical integration formula to compute the normalizing constant of Fisher--Bingham distributions.
This formula uses a numerical integration formula with the continuous Euler transform to a Fourier-type integral representation of the normalizing constant. 
As this method is fast and accurate, it can be applied to the calculation of the normalizing constant of high-dimensional Fisher--Bingham distributions.  
More precisely, the error decays exponentially with an increase in the integration points, and the computation cost increases linearly with the dimensions.
In addition, this formula is useful for calculating the gradient and Hessian matrix of the normalizing constant. 
Therefore, we apply this formula to efficiently calculate the maximum likelihood estimation (MLE) of high-dimensional data.
Finally, we apply the MLE to the hyperspherical variational auto-encoder (S-VAE), a deep-learning-based generative model that restricts the latent space to a unit hypersphere.
We use the S-VAE trained with images of handwritten numbers to estimate the distributions of each label. 
This application is useful for adding new labels to the models. 

\end{abstract}

\section{Introduction}
\label{intro}
\subsection{Fisher--Bingham distribution}
\label{FBd}
The Fisher--Bingham distribution is defined as a multivariate normal distribution restricted on a unit sphere. 

\begin{definition}
	　\\
	For a $p$-dimensional multivariate normal distribution with a mean $\mu$ and a variance-covariance matrix $\Sigma$, the Fisher--Bingham distribution is given by the density function	
	\begin{align*}
	f(x;\mu,\Sigma):=\frac{1}{\mathcal{C}}\exp{\left(-\frac{x^T\Sigma^{-1}x}{2}+x^T\Sigma^{-1}\mu\right)} \mathrm{d}_{\mathcal{S}^{p-1}}(x), \nonumber
	\end{align*}
	where $x \in \mathbb{R}^p$ and
	\begin{align*}
	&\mathcal{C}=\mathcal{C}\left(\frac{\Sigma^{-1}}{2},\Sigma^{-1}\mu\right) \\
	:=&\int_{\mathcal{S}^{p-1}}\exp{\left(-\frac{x^T\Sigma^{-1}x}{2}+x^T\Sigma^{-1}\mu\right)} \mathrm{d}_{\mathcal{S}^{p-1}}(x) \nonumber
	\end{align*}
	is the normalizing constant and 
	$\mathrm{d}_{\mathcal{S}^{p-1}}(x)$ is the uniform measure in the $(p-1)$-dimensional sphere $\mathcal{S}^{p-1}$.
\end{definition}

The Fisher--Bingham distribution plays an essential role in directional statistics, which is concerned with data on various manifolds, especially data represented in a high-dimensional sphere. 
For example, wind direction and the geomagnetic field are common types of data that can be represented on a sphere $\mathcal{S}^2$.
In addition, data on a hypersphere are used in link prediction of networks and image generation.
Therefore, the Fisher--Bingham distribution, a normal distribution restricted on a unit sphere, is commonly used in this field.

However, the spherical domain causes some problems when using Fisher--Bingham distributions. 
One such problems is calculating the normalizing constant. 
As it is difficult to calculate it analytically, a numerical method is necessary. 
The saddlepoint approximation method is a numerical method for computing the normalizing constant $\mathcal{C}(\theta,\gamma)$ developed by \cite{kume2005saddlepoint}. 
Another approach, the holonomic gradient method considered by \cite{kume2018exact}, computes the normalizing constant as well.
However, these methods have some limitations.  
The saddlepoint approximation method is not as accurate as the holonomic gradient method, which is theoretically exact because the problem of calculating $\mathcal{C}(\theta,\gamma)$ is mathematically characterized by solving an ODE. 
However, the holonomic gradient method is computationally expensive and cannot be applied to calculate the normalizing constant of high-dimensional distributions. 
Hence, it is necessary to create a numerical method that is efficient, numerically stable, and accurate.

To construct such a numerical method, the following details about Fisher--Bingham distributions are required (\cite{kume2018exact}). 

Since any orthogonal transformation in $\mathcal{S}^{p-1}$ is isometric, the parameter dimensions are reduced from $(p \times p+p)$ to $2p$ by singular value decomposition. 
Therefore, we have
\begin{align*}
\mathcal{C}\left(\frac{\Sigma^{-1}}{2},\Sigma^{-1}\mu\right)=\mathcal{C}\left(\frac{\Delta^{-1}}{2},\Delta^{-1}\text{O}\mu\right), 
\end{align*}
where $\Delta=\text{diag}(\delta_1^2,\cdots,\delta_p^2) $ and $O$ is the orthogonal matrix obtained from $\Sigma =\text{O}^T\Delta \text{O}$. 
Thus, without loss of generality, we can assume that the variance-covariance matrix $\Sigma$ is diagonal. 
After reducing the parameter dimensions to $2p$, the normalizing constant becomes
\begin{align*}
&\mathcal{C}\left(\frac{\Delta^{-1}}{2},\Delta^{-1}\text{O}\mu\right)=\mathcal{C}(\theta,\gamma)  \\
:=&\int_{\mathcal{S}^{p-1}} \exp\left({\sum_{i=1}^{p}(-\theta_i x_i^2 + \gamma_i x_i )}\right) \mathrm{d}_{\mathcal{S}^{p-1}}(x),
\end{align*}
where 
\begin{align*}
\theta=(\theta_1,\cdots,\theta_p)=\left(\frac{1}{2\delta_1^2},\cdots,\frac{1}{2\delta_p^2}\right)=\text{diag}\left(\frac{\Delta^{-1}}{2}\right) 
\end{align*}
and
\begin{align*}
\gamma=(\gamma_1,\cdots,\gamma_p)=\Delta^{-1}\text{O}\mu. 
\end{align*}
Since $x$ is restricted on a unit sphere, we have
\begin{align*}
&\mathcal{C}(\theta+cI,\gamma) \\
=&\int_{\mathcal{S}^{p-1}} \exp\left({\sum_{i=1}^{p}(-(\theta_i+c) x_i^2 + \gamma_i x_i )}\right) \mathrm{d}_{\mathcal{S}^{p-1}}(x) \\
=&\int_{\mathcal{S}^{p-1}} \exp \left(-c+\left({\sum_{i=1}^{p}(-\theta_i x_i^2 + \gamma_i x_i )}\right) \right)\mathrm{d}_{\mathcal{S}^{p-1}}(x) \\
=&e^{-c} \int_{\mathcal{S}^{p-1}}\exp\left({\sum_{i=1}^{p}(-\theta_i x_i^2 + \gamma_i x_i )}\right) \mathrm{d}_{\mathcal{S}^{p-1}}(x) \\
=&e^{-c}\mathcal{C}(\theta,\gamma),
\end{align*}
where $c$ is a real number and $I=(1,1,\cdots,1) \in \mathbb{R}^p$. 
If we put
\begin{align*}
f(x;\theta,\gamma):=\frac{1}{\mathcal{C}(\theta,\gamma)}\exp \left( \sum_{i=1}^{p}(-\theta_i x_i^2+\gamma_i x_i) \right)\mathrm{d}_{\mathcal{S}^{p-1}}(x).
\end{align*}
then we have
\begin{align*}
&f(x;\theta+cI,\gamma) \\
=&\frac{1}{\mathcal{C}(\theta+cI,\gamma)}\exp \left( \sum_{i=1}^{p}(-(\theta_i+c) x_i^2+\gamma_i x_i) \right)\mathrm{d}_{\mathcal{S}^{p-1}}(x) \\
=&\frac{e^c}{\mathcal{C}(\theta,\gamma)} \exp \left(-c+\left( \sum_{i=1}^{p}(-\theta_i x_i^2+\gamma_i x_i) \right)\right)\mathrm{d}_{\mathcal{S}^{p-1}}(x) \\
=&\frac{1}{\mathcal{C}(\theta,\gamma)}\exp \left( \sum_{i=1}^{p}(-\theta_i x_i^2+\gamma_i x_i) \right)\mathrm{d}_{\mathcal{S}^{p-1}}(x) \\
=&f(x;\theta,\gamma).
\end{align*}
As a result, if the normalizing constant $\mathcal{C}(\theta,\gamma)$ is obtained, $\mathcal{C}(\theta+cI,\gamma)$ can also be obtained. 
Moreover, for the maximum likelihood estimation (MLE), as $f(x;\theta,\gamma)=f(x;\theta+cI,\gamma)$, $\theta$ can be shifted to $\theta+cI$ for all $c \in \mathbb{R}$.
Additionally, because the unit sphere is symmetrical,
\begin{align*}
\mathcal{C}(\theta,|\gamma|)&=\int_{\mathcal{S}^{p-1}} \exp\left({\sum_{i=1}^{p}(-\theta_i x_i^2 + |\gamma_i| x_i )}\right) \mathrm{d}_{\mathcal{S}^{p-1}}(x) \\
&=\int_{\mathcal{S}^{p-1}} \exp\left({\sum_{i=1}^{p}(-\theta_i x_i^2 + \gamma_i x_i )}\right) \mathrm{d}_{\mathcal{S}^{p-1}}(x) \\
&=\mathcal{C}(\theta,\gamma).
\end{align*}
As a result, it can be assumed that $\gamma$ has non-negative entries when calculating the normalizing constant.

\subsection{Aim of this paper}
\label{aim}
In this paper, 
\begin{enumerate}
	\item we propose an efficient numerical integration formula to compute the normalizing constant.
	\item we apply this formula to perform MLE.
	\item we apply MLE to the latent variables of the hyperspherical variational auto-encoder (S-VAE) (\cite{davidson2018hyperspherical}).
\end{enumerate}

The normalizing constant of Fisher--Bingham distributions can be represented in a Fourier integration form.
Therefore, we can use the numerical integration formula with the continuous Euler transform introduced by \cite{ooura2001continuous}. 
Note that the continuous Euler transform is useful for calculating the normalizing constant and MLE.

This method can be applied to the MLE of high-dimensional data, such as the latent variables of S-VAE (\cite{davidson2018hyperspherical}), a generating model used in machine learning. 
The dimensions of the hyperspherical variational auto-encoder rely on the complexity of the data. 
For example, for human face data, there may be 100 dimensions of the latent variables.   

\subsection{Organization of this paper}
\label{organization}
This paper is organized as follows. 
In Section 2, we make some general remarks about the Fisher--Bingham distribution and the Fourier transform representation of the normalizing constant. 
In Section 3, we explain the continuous Euler transform and its use for numerical computation of the normalizing constant. 
In Section 4, we discuss the calculation of the gradient of the normalizing constant, which is necessary for MLE. 
Subsequently, the MLE algorithm is provided. 
In Section 5, we demonstrate some MLE numerical experiment to show the effectiveness of this method.
In Section 6, we show the application of MLE in the S-VAE whose latent space includes high-dimensional data on a hypersphere.  

\section{Fourier transform representation of the normalizing constant}
\label{Ft nc}
\subsection{Laplace inversion representation}
\label{Laplace}
This section explains how the normalizing constant can be represented in a simpler form, as derived by \cite{kume2005saddlepoint} and \cite{kume2018exact}. 
The first step is to change the range of integration of the normalizing constant from  a $(p-1)$-dimensional hypersphere $\mathcal{S}^{p-1}$ to a one-dimensional line $i\mathbb{R}+t_0$. 
The integration range is then shifted to $\mathbb{R}$. 
Note that the derivation can be somewhat technical, but the calculation itself is not difficult.

First, the distribution $f$ of $p$ independent normal random variables $X_i \sim \mathcal{N}(\mu_i,\frac{1}{2\theta_i}) (i=1,\cdots,p)$ is
\begin{align*}
f(x_1,\cdots,x_p)=\frac{\prod_{i=1}^p\theta_i^{\frac{1}{2}}}{\pi^{\frac{p}{2}}}\exp \left(-\sum_{i=1}^p \theta_i(x_i-\mu_i)^2\right).
\end{align*}
We then apply the variable transform
\begin{align*}
\begin{cases}
&r=\sum_{i=1}^p x_i^2=x^T x \\
&\phi=(\phi_1,\cdots,\phi_p)=\left(\frac{x_1}{r^{\frac{1}{2}}},\cdots,\frac{x_p}{r^{\frac{1}{2}}}\right)=\frac{x}{r^{\frac{1}{2}}} 
\end{cases}
\end{align*}
to $f(x_1,\cdots,x_p)$ and integrate it with respect to $\phi$. Then, the marginalized distribution becomes
\begin{align}
f_{\text{mrg}}(r)=&\frac{1}{2}\pi^{-\frac{p}{2}}\left(\prod_{i=1}^p \theta_i^{\frac{1}{2}}\right) \hat{\mathcal{C}}(r\theta,r^{\frac{1}{2}}\gamma) \nonumber \\
&\times \exp\left(-\frac{1}{4}\sum_{i=1}^p \frac{\gamma_i^2}{\theta_i}\right)r^{\frac{p}{2}-1},
\label{eq1}
\end{align}
where
\begin{align*}
\gamma=(2\theta_1 \mu_1,\cdots,2\theta_p \mu_p)
\end{align*} 
and
\begin{align}
&\hat{\mathcal{C}}(r\theta,r^{\frac{1}{2}}\gamma) \nonumber \\
=&\int_{\mathcal{S}^{p-1}}\exp\left(-\sum_{i=1}^p(r\theta_i \phi_i^2-r^{\frac{1}{2}}\gamma_i\phi_i)\right)\mathrm{d}_{\mathcal{S}^{p-1}}(\phi).
\label{eq2}
\end{align}
When $r=1$, Equation (\ref{eq2}) matches the definition of the normalizing constant $\mathcal{C}(\theta,\gamma)$. As a result, based on Equation (\ref{eq1}), we obtain
\begin{align}
\mathcal{C}(\theta,\gamma)=2\pi^{\frac{p}{2}}\left(\prod_{i=1}^p \theta_i^{-\frac{1}{2}}\right)f_{\text{mrg}}(1)\exp\left(\frac{1}{4}\sum_{i=1}^p\frac{\gamma_i^2}{\theta_i}\right). 
\label{eq3}
\end{align}
Therefore, if the distribution $f_{\text{mrg}}(r)$ can be represented in a one-dimensional integration form, the goal will be achieved.

The moment generating function of $f_{\text{mrg}}(r)$ is
\begin{align*}
\mathcal{L}(t)=\frac{\exp\left(\sum_{i=1}^p\left(\frac{\gamma_i^2}{4(\theta_i+t)}-\frac{\gamma_i^2}{4\theta_i}\right)\right)}{\prod_{i=1}^p\sqrt{1+\frac{t}{\theta_i}}}. 
\end{align*} 
Since $r=x^T x \geq 0$, the moment generating function $\mathcal{L}(t)$ is the same as the Laplace transform of $f_{\text{mrg}}(r)$. Thus, with the inverse Laplace transform, we obtain
\begin{align}
&f_{\text{mrg}}(r)=\frac{1}{2\pi i}\int_{i\mathbb{R}+t_0}\mathcal{L}(t)e^{rt}\mathrm{d}t,
\label{eq4}
\end{align}
where $t_0 \geq 0$.
Substituting Equation (\ref{eq4}) into Equation (\ref{eq3}), we get
\begin{align}
&\mathcal{C}(\theta,\gamma)=-i\pi^{\frac{p}{2}-1}\int_{i\mathbb{R}+t_0}\prod_{i=1}^p\frac{\exp\left(\frac{\gamma_i^2}{4(\theta_i+t)}\right)}{\sqrt{\theta_i+t}}e^t \mathrm{d}t.
\label{eq5}
\end{align}

This is the Laplace inversion representation of the normalizing constant, which contains a one-parameter integration of $i\mathbb{R}+t_0$. 
The Fourier transform representation can be easily deduced by applying the variable transform to Equation (\ref{eq5}). 

\subsection{Fourier-type integral representation}
\label{Fourier}
\begin{thm}
	\begin{align}
	\mathcal{C}(\theta,\gamma)=\pi^{\frac{p}{2}-1}e^{-t_0}\int_{\mathbb{R}} \hspace{0.2cm} \mathcal{A}(t;\theta,\gamma) e^{it} \mathrm{d}t,  
	\label{eq01}
	\end{align}
	where
	\begin{align*}
	\mathcal{A}(t;\theta,\gamma)=\prod_{i=1}^p\frac{\exp\left(\frac{\gamma_i^2}{4(\theta_i-it-t_0)}\right)}{\sqrt{\theta_i-it-t_0}}.
	\end{align*}
\end{thm}

Now, we derive the representation of the normalizing constant in the form of a Fourier integral.
Therefore, the next step is to apply a numerical integration formula to calculate this Fourier integral.

\section{Continuous Euler transform for numerical calculation}
\label{CE nc}
\subsection{Continuous Euler transform}
\label{Continuous}
Since the normalizing constant is represented as a Fourier-type integral in Equation (\ref{eq01}),
it is necessary to find an accurate numerical integration formula to calculate this integration.
Here, note that the function $\mathcal{A}(t;\theta,\gamma)$ decays slowly.
With a usual trapezoidal formula, the slower the function decays, the slower the convergence of the numerical integration becomes.
It is necessary to reinforce the decay of the function $\mathcal{A}(t;\theta,\gamma)$ to improve the accuracy.
Here, we can adopt the continuous Euler transform by \cite{ooura2001continuous}, which can accelerate the convergence of the Fourier transform.
Moreover, \cite{tanaka2014error} proved that adding the continuous Euler transformation to the trapezoidal formula made the integration converge rapidly. 
The details of the continuous Euler transform have been provided by \cite{ooura2001continuous} and \cite{tanaka2014error}. 

We apply the continuous Euler transform to Equation (\ref{eq01}).
Choose $\omega_d$ and $\omega_u$ to satisfy $\omega_d \leq 1\leq \omega_u$ and $\omega_d/\omega_u\leq1/2$. Choose $d \leq \min\{ |\theta_i-t_0| \}$.
Let $N$ be an integer with
\begin{align*}
N \geq \frac{2d(\omega_d+\omega_u)\omega_u^2}{\pi \omega_d^2}.
\end{align*}
Let $h,p$, and $q$ be defined by 
\begin{align*}
h=\sqrt{\frac{2\pi d(\omega_d+\omega_u)}{\omega_d^2 N}},
\hspace{0.1cm}
p=\sqrt{\frac{Nh}{\omega_d}},
\hspace{0.1cm}
q=\sqrt{\frac{\omega_d N h}{4}}.
\end{align*}
Then, we get
\begin{align}
	&\mathcal{C}(\theta,\gamma) \nonumber \\
	=&\pi^{\frac{p}{2}-1}e^{-t_0}\int_{\mathbb{R}} \hspace{0.2cm} \mathcal{A}(t;\theta,\gamma) e^{it} \mathrm{d}t \nonumber \\
	\approx& \pi^{\frac{p}{2}-1}e^{-t_0}\int_{\mathbb{R}} \hspace{0.2cm} w(|t|,p,q)\mathcal{A}(t;\theta,\gamma) e^{it} \mathrm{d}t \\
	\approx& \pi^{\frac{p}{2}-1}e^{-t_0} h\sum_{n=-N-1}^{N}w(|nh|;p,q)\mathcal{A}(nh,\theta,\gamma)e^{inh} \label{eq02} \\
	=:& \mathcal{C}_w^{(N,h)}(\theta,\gamma),
\end{align}
where
\begin{align*}
&w(x;p,q)=\frac{1}{2}\text{erfc}\left(\frac{x}{p}-q\right).
\end{align*}
Equation (\ref{eq02}) uses the trapezoidal formula.

The accuracy of the numerical formula $\mathcal{C}_w^{(N,h)}(\theta,\gamma)$ is prooved using Theorem \ref{thm2}.

\begin{thm}
	\begin{align*}
	|\mathcal{C}(\theta,\gamma)-\mathcal{C}_{w}^{(N,h)}(\theta,\gamma)| \leq \text{Poly}(N)\exp\left( -\sqrt{\frac{\pi d \omega_d^2 N}{2(\omega_d+\omega_u)}} \right),
	\end{align*}
	where $\text{Poly}(N)$ is a polynomial of $N$.
	\label{thm2}
\end{thm}

With this theorem, if the normalizing constant $\mathcal{C}(\theta,\gamma)$ is approximated by $\mathcal{C}_{w}^{(N,h)}(\theta,\gamma)$, the error converges to $0$ as $\sqrt{N}\rightarrow \infty$.
In other words, any accuracy can be achieved if a practically large enough $N$ is taken.
For instance, if the normalizing constant with an error less than $10^{-6}$ is necessary, $N=200$ is sufficient.
Subsequently, if the parameter dimensions are about 100, it only takes 20 ms, whereas the holonomic gradient method takes about 15 s with 10 dimensions (\cite{sei2015calculating}). 
The details of the numerical experiments are provided in Section 5.
An efficient numerical integration formula for the normalizing constant can be  obtained using this method. 

\section{MLE optimization using the continuous Euler transform}
\label{MLE CE}
The likelihood function of the Fisher--Bingham distribution has been provided by  \cite{kume2018exact}. With a observed data matrix $X=(x_1,x_2,...,x_n) \in \mathbb{R}^{p \times n}$, put $A=\frac{\sum_{i=1}^{n}x_i x_i^T}{n}$ and $B=\frac{\sum_{i=1}^{n}x_i}{n}$, the likelihood function is 
\begin{align*}
&\log \mathcal{L}(\frac{\Sigma^{-1}}{2},\Sigma^{-1}\mu,X) \\
=&\log \prod_{i=1}^{n}\left( \frac{\exp{\left(x_i^T\Sigma^{-1}\mu-\frac{x_i^T\Sigma^{-1}x_i}{2}\right)} }{\mathcal{C}\left(\frac{\Sigma^{-1}}{2},\Sigma^{-1}\mu\right)}\right) \\
=&-n\log \mathcal{C}\left(\frac{\Delta^{-1}}{2},\gamma \right)-\sum_{i=1}^{n}\left( x_i^T\frac{\Sigma^{-1}}{2} x_i -x_i^T\Sigma^{-1}\mu \right)\\
=&-n\log \mathcal{C}\left(\frac{\Delta^{-1}}{2},\gamma \right)-n\text{tr}\left( AO^T\frac{\Delta^{-1}}{2}O-OB\gamma^T \right) \\
=&-n\left( \log \mathcal{C}(\theta,\gamma)+\text{tr}(AO^T\text{diag}(\theta)O+OB\gamma^T) \right)
\end{align*}
with
\begin{align*}
&\Sigma^{-1}=O^T\Delta^{-1}O, \\
&\hspace{0.1cm}
\mathcal{C}\left(\frac{\Sigma^{-1}}{2},\Sigma^{-1}\mu\right) 
=\mathcal{C}\left(\frac{\Delta^{-1}}{2},\gamma\right)=\mathcal{C}(\theta,\gamma), \hspace{0.1cm} \\
&\gamma=\Delta^{-1}O\mu 
\end{align*} 
and
\begin{align*}
\frac{\Delta^{-1}}{2}=\text{diag}(\theta).
\end{align*}
Therefore, maximizing the likelihood function 
\begin{align*}
\log \mathcal{L}\left(\frac{\Sigma^{-1}}{2},\Sigma^{-1}\mu,X\right)
\end{align*}
is equivalent to minimizing
\begin{align}
&\log L (\theta,\gamma,O) \nonumber \\
:=&\log \mathcal{C}(\theta,\gamma)+\text{tr}(AO^T\text{diag}(\theta)O+OB\gamma^T).
\label{eq8}
\end{align}
In this section, $\log L(\theta,\gamma,O)$ is also called the likelihood function, although
\begin{align*}
\log L(\theta,\gamma,O)=-\frac{1}{n} \log \mathcal{L}\left(\frac{\Sigma^{-1}}{2},\Sigma^{-1}\mu,X\right).
\end{align*}

Thus, it is possible to optimize Equation (\ref{eq8}) by iteratively updating the parameters that decrease the likelihood value. 
First, we consider the optimization problem in $\theta$ and $\gamma$ for a fixed $O$. Then, the optimal $O$ can be obtained by minimizing $\text{tr}(AO^T\text{diag}(\theta)O+OB\gamma^T)$ mantaining the value of $\theta$ and $\gamma$ fixed,.
The formula for updating $O$ has been published by \cite{kume2018exact}. 
Although the outline for updating $O$ is mentioned in this section, $O$ is fixed as an identity matrix, that is, $\Sigma$ is assumed to be diagonal in numerical experiments.  

It is necessary to calculate the partial derivatives of $\log L(\theta,\gamma,O)$ when optimizing $\theta$ and $\gamma$.
\begin{align}
\frac{\partial \log L(\theta,\gamma,O)}{\partial \theta}=\frac{\partial \mathcal{C}(\theta,\gamma)}{\partial \theta} \frac{1}{\mathcal{C}(\theta,\gamma)}+\text{diag}(OAO^T)
\label{eq9}
\end{align}
and
\begin{align}
\frac{\partial \log L(\theta,\gamma,O)}{\partial \gamma}=\frac{\partial \mathcal{C}(\theta,\gamma)}{\partial \gamma} \frac{1}{\mathcal{C}(\theta,\gamma)}+B^T O^T.
\label{eq10}
\end{align}
The partial derivatives of $\mathcal{C}(\theta,\gamma)$ are needed to calculate Equation (\ref{eq9}) and Equation (\ref{eq10}).

In Theorem 1, the Fourier transform representation of the normalizing constant becomes
\begin{align*}
\mathcal{C}(\theta,\gamma)=\pi^{\frac{p}{2}-1}e^{t_0}\int_{\mathbb{R}} \hspace{0.2cm} \mathcal{A}(t;\theta,\gamma) e^{it} \mathrm{d}t,
\end{align*}
where
\begin{align*}
\mathcal{A}(t;\theta,\gamma)=\prod_{i=1}^p\frac{\exp\left(\frac{\gamma_i^2}{4(\theta_i-it-t_0)}\right)}{\sqrt{\theta_i-it-t_0}}.
\end{align*}
By changing the order of integration and differentiation, the partial  derivatives of $\mathcal{C}(\theta,\gamma)$ become
\begin{align*}
\mathcal{C}_{\theta_i}(\theta,\gamma)&:=\frac{\partial \mathcal{C}(\theta,\gamma)}{\partial \theta_i} \\
&=\pi^{\frac{p}{2}-1}e^{t_0}\int_{\mathbb{R}} \hspace{0.2cm} \frac{\partial \mathcal{A}(t;\theta,\gamma)}{\partial \theta_i} e^{it} \mathrm{d}t, \\
\mathcal{C}_{\gamma_i}(\theta,\gamma)&:=\frac{\partial \mathcal{C}(\theta,\gamma)}{\partial \gamma_i} \\
&=\pi^{\frac{p}{2}-1}e^{t_0}\int_{\mathbb{R}} \hspace{0.2cm} \frac{\partial \mathcal{A}(t;\theta,\gamma)}{\partial \gamma_i} e^{it} \mathrm{d}t.
\end{align*}

The continuous Euler transform can also be applied to these calculations to numerically calculate the derivatives efficiently.  

	If we put
	\begin{align*}
	&\mathcal{A}_{\theta_i}(t;\theta,\gamma):=\frac{\partial \mathcal{A}(t;\theta,\gamma)}{\partial \theta_i} \nonumber \\ 
	=&\left\{ \exp \left( \frac{\gamma_i^2}{4(\theta_i-it-t_0)}\right) \frac{1}{2(\theta_i-it-t_0)} \right. \\ 
	&\times \left. \left( -\frac{\gamma_i^2}{2(\theta_i-it-t_0)}-1 \right)  \right\} \prod_{j \neq i}\frac{\exp \left( \frac{\gamma_i^2}{4(\theta_i-it-t_0)} \right)}{\sqrt{\theta_i-it-t_0}}, \\
	&\mathcal{A}_{\gamma_i}(t;\theta,\gamma):=\frac{\partial \mathcal{A}(t;\theta,\gamma)}{\partial \gamma_i} \nonumber \\
	=&\frac{\exp \left( \frac{\gamma_i^2}{4(\theta_i-it-t_0)} \right)\gamma_i}{2(\theta_i-it-t_0)^{\frac{3}{2}}}\prod_{j \neq i}\frac{\exp \left( \frac{\gamma_i^2}{4(\theta_i-it-t_0)} \right)}{\sqrt{\theta_i-it-t_0}}.
	\end{align*}	
	Then, we get
	\begin{align}
	&\mathcal{C}_{\theta_i}(\theta,\gamma) \nonumber \\
	=&\pi^{\frac{p}{2}-1}e^{t_0}\int_{\mathbb{R}} \mathcal{A}_{\theta_i}(t;\theta,\gamma)e^{it} \mathrm{d}t \nonumber\\
	\approx&\pi^{\frac{p}{2}-1}e^{t_0} h\sum_{n=-N-1}^{N}w(|nh|;p,q)\mathcal{A}_{\theta_i}(nh,\theta,\gamma)e^{inh}, 
	\label{eq11}
	\end{align}
	\begin{align}
	&\mathcal{C}_{\gamma_i}(\theta,\gamma) \nonumber \\
	=&\pi^{\frac{p}{2}-1}e^{t_0}\int_{\mathbb{R}} \hspace{0.2cm}  \mathcal{A}_{\gamma_i}(t;\theta,\gamma)e^{it} \mathrm{d}t \nonumber \\
	\approx& \pi^{\frac{p}{2}-1}e^{t_0} h\sum_{n=-N-1}^{N}w(|nh|;p,q)\mathcal{A}_{\gamma_i}(nh,\theta,\gamma)e^{inh}.
	\label{eq12}
	\end{align}

Therefore, the derivatives of the normalizing constant, as well as the derivatives of the likelihood function, can be calculated using Equation (\ref{eq9}) and (\ref{eq10}). 
As a result, it is possible to optimize $\theta$ and $\gamma$ with a fixed $O$.

Given that the algorithm to optimize $O$ with fixed $\theta$ and $\gamma$ values was developed by \cite{kume2018exact}, this will not be demonstrated in this paper.   

\begin{align}
\mathcal{A}=\text{diag}(\theta)OAO^T-OAO^T\text{diag}(\theta)+\gamma B^TO^T
\label{eq13}
\end{align} 
and
\begin{align}
\hat{v}=\mathcal{A}-\mathcal{A}^T.
\label{eq14}
\end{align}

$\mathcal{A}$ must be symmetrical for $O$ to be an optional orthogonal matrix (\cite{kume2018exact}). Moreover, if $\mathcal{A}$ is symmetrical, a curve $Oe^{\hat{v}t}$ with $t$ reduces to a single point; this can be used as a stopping criterion. 
The proof of this stopping criterion was provided by \cite{kume2018exact}.  

Now, we have obtained all parts necessary to optimize the likelihood function.
Consequently, we can discuss the approach for obtaining the MLE of Fisher--Bingham distributions. 
This algorithm performs the same steps as the algorithm given by  \cite{kume2018exact}, but the method used to calculate the derivatives of the likelihood function and the normalizing constant are different. 
Although Kume and Sei used the holonomic gradient method, the continuous Euler transform is adopted in this paper.

\begin{algorithm}{(the gradient descent method)\\}
	Update the given $\theta$, $\gamma$, and $O$ as follows until the differentiation of $\theta$, $\gamma$, and $\hat{v}$ becomes small enough:
	\begin{enumerate}
		\item \begin{align*} \hat{\theta}=\theta+\frac{\partial \log L(\theta,\gamma,O)}{\partial \theta} \delta_{\theta}.
		\end{align*}
		The partial derivatives of the likelihood function $\displaystyle \frac{\partial \log L(\theta,\gamma,O)}{\partial \theta}$ are obtained by substituting Equation (\ref{eq11}) into Equation (\ref{eq9}), and $\delta_{\theta}$ is a real number such that 
		\begin{align*}
		\log L(\hat{\theta},\gamma,O)<\log L(\theta,\gamma,O).
		\end{align*}
		\item \begin{align*}
			\hat{\gamma}=\gamma+\frac{\partial \log L(\theta,\gamma,O)}{\partial \gamma} \delta_{\gamma}.
		\end{align*}
		The partial derivatives of the likelihood function $\displaystyle \frac{\partial \log L(\theta,\gamma,O)}{\partial \gamma}$ are obtained by substituting Equation (\ref{eq12}) into Equation (\ref{eq10}), and $\delta_{\gamma}$ is a real number such that 
		\begin{align*}
		\log L(\theta,\hat{\gamma},O)<\log L(\theta,\gamma,O).
		\end{align*}
		\item \begin{align*}\hat{O}=e^{\hat{v}t_0}O.
		\end{align*}
		$\hat{v}$ is obtained by substituting Equation (\ref{eq13}) into Equation (\ref{eq14}), and $t_0$ is a real number such that 
		\begin{align*}
		\log L(\theta,\gamma,\hat{O})<\log L(\theta,\gamma,O).
		\end{align*}
	\end{enumerate}
\end{algorithm}

This algorithm is based on the gradient descent method that converges linearly. If faster convergence is required, it is preferable to use the quasi-Newton method. 

\section{Numerical experiments}
\label{experiments}
\subsection{Normalizing constant}
\label{nc}
We compare the calculation method for the normalizing constant of the Bingham distribution, the holonomic gradient method, and the saddlepoint approximation method with our method.
The Bingham distribution is a special case of the Fisher--Bingham distribution, which fixes $\gamma=0$. 

\begin{definition}
	For a $p$-dimensional multivariate distribution, the Bingham distribution is	given by the density function
	\begin{align*}
	f(x;\theta):=\frac{1}{\mathcal{C}(\theta)}\exp{\left(-\sum_{i=1}^{p}\theta_i x_i^2\right)} \mathrm{d}_{\mathcal{S}^{p-1}}(x),
	\end{align*}
	where $x \in \mathbb{R}^p$ and
	\begin{align*}
	\mathcal{C}\left(\theta  \right):=\int_{\mathcal{S}^{p-1}}\exp{\left(-\sum_{i=1}^{p}\theta_i x_i^2\right)} \mathrm{d}_{\mathcal{S}^{p-1}}(x)
	\end{align*}
	is the normalizing constant and 
	$\mathrm{d}_{\mathcal{S}^{p-1}}(x)$ is the uniform measure in the $(p-1)$-dimensional sphere $\mathcal{S}^{p-1}$.
\end{definition}

Theorem 2 is used to compute the normalizing constant of the Bingham distribution.
More precisely, the numerical integration formula is 
\begin{align*}
\mathcal{C}_w^{(N,h)}(\theta)=\pi^{\frac{p}{2}-1}e^{t_0} h\sum_{n=-N-1}^{N}w(|nh|;p,q)\mathcal{A}(nh,\theta)e^{inh},
\end{align*}
where
\begin{align*}
\mathcal{A}(t;\theta)=\prod_{i=1}^p\frac{1}{\sqrt{\theta_i-it-t_0}}.
\end{align*}

The same parameters were used by \cite{sei2015calculating}. 
The results of the holonomic gradient method and the saddlepoint approximation method were obtained from \cite{sei2015calculating}.
In our method, the parameter $N$ is fixed to 200. 
The holonomic gradient method must be theoretically exact because the problem is mathematically characterized by solving an ODE with numerical methods.

The accuracy of the holonomic gradient method and our method is very high. 
As shown in Table \ref{the normalizing constant} to Table \ref{the normalizing constant4}, the normalizing constants with different parameters calculated by these methods (the hg and ce columns) coincide until they reach 6 decimal places. 
However, the saddlepoint approximation method is not as good because the error is larger than $10^{-3}$. 
Besides, the calculation time for our method with 100 dimensions is about 20 ms, whereas the holonomic gradient method takes about 15 s with 10 dimensions (\cite{sei2015calculating}).

In Table \ref{the normalizing constant}, columns 2 and 3 compare the saddlepoint approximation (spa) and the holonomic gradient method (hg) of $\theta=(0,1,2,\kappa)$ with our method (ce); 
columns 4 and 5 compare the same quantities for $\theta=(0,1,2,\kappa,\kappa)$.

\begin{table*}[htb]
\caption{The normalizing constant of Bingham distributions with $\theta=(0,-1,-2,\kappa)$ in columns 2, 3, and 4 and $\theta=(0,-1,-2,-\kappa,-\kappa)$ in columns 5, 6, and 7}
	\label{the normalizing constant}
	\centering
		\begin{tabular}{ccccccc}\hline
			$\kappa$ & spa      & hg       & ce       & spa      & hg       & ce \rule[-2mm]{0mm}{6mm} \\ \hline
			5        & 4.237006 & 4.238950 & 4.238950 & 3.376766 & 3.372017 & 3.372017 \rule[-2mm]{0mm}{7mm} \\
			10       & 2.982628 & 2.985576 & 2.985576 & 1.689684 & 1.689355 & 1.689355 \rule[-2mm]{0mm}{6mm} \\
			30       & 1.708766 & 1.711919 & 1.711919 & 0.555494 & 0.556123 & 0.556123  \rule[-2mm]{0mm}{6mm} \\
			50       & 1.321178 & 1.323994 & 1.323994 & 0.332102 & 0.332661 & 0.332661  \rule[-2mm]{0mm}{6mm} \\
			100      & 0.932895 & 0.935094 & 0.935094 & 0.165587 & 0.165940 & 0.165940  \rule[-2mm]{0mm}{6mm} \\
			200      & 0.659185 & 0.660814 & 0.660814 & 0.082676 & 0.082871 & 0.082871 \rule[-3mm]{0mm}{7mm} \\ \hline
		\end{tabular}
\end{table*}

The complex Bingham distribution, which is defined on a unit complex sphere, can be calculated analytically (\cite{kume2005saddlepoint}).

\begin{definition}
	For a $p$-dimensional multivariate distribution, the complex Bingham distribution is	given by the density function
	\begin{align*}
	f(x;\theta):=\frac{1}{\mathcal{C}(\theta)}\exp{\left(-\sum_{i=1}^{p}\theta_i x_i^2\right)} \mathrm{d}_{\mathcal{S}^{p-1}}(x),
	\end{align*}
	where $x \in \mathbb{C}^p$ and
	\begin{align*}
	\mathcal{C}\left(\theta  \right):=\int_{\mathcal{S}^{p-1}}\exp{\left(-\sum_{i=1}^{p}\theta_i x_i^2\right)} \mathrm{d}_{\mathcal{S}^{p-1}}(x)
	\end{align*}
	is the normalizing constant and 
	$\mathrm{d}_{\mathcal{S}^{p-1}}(x)$ is the uniform measure in the $(p-1)$-dimensional complex sphere $\mathcal{S}^{p-1}$.
\end{definition} 

In Table \ref{the normalizing constant2}, we compare the saddlepoint, exact, hg, and our method for the complex Bingham distribution with parameters $\phi=(0,1,2,\kappa)$, that is, $\theta=(0,0,1,2,\kappa,\kappa)$. 

\begin{table}[htb]
		\caption{The normalizing constant of complex Bingham distributions with $\theta=(0,-1,-2,-\kappa)$} 
		\label{the normalizing constant2}
		\centering
		\begin{tabular}{ccccc} \hline
			$\kappa$ & spa      & ex       & hg       & ce \rule[-2mm]{0mm}{6mm}\\ \hline
			5        & 5.942975 & 5.936835 & 5.936835 & 5.936835 \rule[-2mm]{0mm}{7mm}\\
			10       & 3.429004 & 3.425468 & 3.425468 & 3.425468 \rule[-2mm]{0mm}{6mm}\\
			30       & 1.248280 & 1.246421 & 1.246421 & 1.246421 \rule[-2mm]{0mm}{6mm}\\
			50       & 0.761347 & 0.760180 & 0.760180 & 0.760180 \rule[-2mm]{0mm}{6mm}\\
			100      & 0.385272 & 0.384675 & 0.384675 & 0.384675 \rule[-2mm]{0mm}{6mm}\\
			200      & 0.193779 & 0.193477 & 0.193477 & 0.193477 \rule[-3mm]{0mm}{7mm}\\ \hline
		\end{tabular}
\end{table}

In Table \ref{the normalizing constant3}, columns 2 and 3 compare the saddlepoint approximation (spa) and the holonomic gradient method (hg) of $\theta=(0,1,22,\kappa)$ with our method (ce); 
columns 4 and 5 compare the same quantities for $\theta=(0,1,22,\kappa,\kappa)$.

\begin{table*}[htb]
\caption{The normalizing constant of Bingham distributions with $\theta=(0,-1,-22,\kappa)$ in columns 2, 3, and 4 and $\theta=(0,-1,-22,-\kappa,-\kappa)$ in columns 5, 6, and 7}
	\label{the normalizing constant3}
	\centering
		\begin{tabular}{ccccccc}\hline
			$\kappa$ & spa      & hg       & ce       & spa      & hg       & ce \rule[-2mm]{0mm}{6mm} \\ \hline
			5        & 1.258672 & 1.273161 & 1.273161 & 1.032128 & 1.044072 & 1.044072 \rule[-2mm]{0mm}{7mm} \\
			10       & 0.874523 & 0.883394 & 0.883394 & 0.500707 & 0.505223 & 0.505223 \rule[-2mm]{0mm}{6mm} \\
			30       & 0.497757 & 0.503213 & 0.503213 & 0.162251 & 0.163901 & 0.163901  \rule[-2mm]{0mm}{6mm} \\
			50       & 0.384440 & 0.388775 & 0.388775 & 0.096784 & 0.097828 & 0.097828  \rule[-2mm]{0mm}{6mm} \\
			100      & 0.271249 & 0.274375 & 0.274375 & 0.048182 & 0.048725 & 0.048725  \rule[-2mm]{0mm}{6mm} \\
			200      & 0.191595 & 0.193826 & 0.193826 & 0.024039 & 0.024316 & 0.024316 \rule[-3mm]{0mm}{7mm} \\ \hline
		\end{tabular}
\end{table*}

In Table \ref{the normalizing constant4}, we compare the saddlepoint, exact, hg, and our method for the complex Bingham distribution with parameters $\phi=(0,1,22,\kappa)$, that is, $\theta=(0,0,1,1,22,22,\kappa,\kappa)$. 

\begin{table}[htb]
		\caption{The normalizing constant of complex Bingham distributions with $\theta=(0,-1,-22,-\kappa)$} 
\label{the normalizing constant4}
\centering
		\begin{tabular}{ccccc} \hline
			$\kappa$ & spa      & ex       & hg & ce \rule[-2mm]{0mm}{6mm}\\ \hline
			5        & 0.921027 & 0.921726 & 0.921726 & 0.921726 \rule[-2mm]{0mm}{7mm}\\
			10       & 0.506236 & 0.506341 & 0.506341 & 0.506341 \rule[-2mm]{0mm}{6mm}\\
			30       & 0.177602 & 0.177495 & 0.177495 & 0.177495 \rule[-2mm]{0mm}{6mm}\\
			50       & 0.177602 & 0.107458 & 0.107458 & 0.107458 \rule[-2mm]{0mm}{6mm}\\
			100      & 0.054115 & 0.054081 & 0.054081 & 0.054081 \rule[-2mm]{0mm}{6mm}\\
			200      & 0.027144 & 0.027127 & 0.027127 & 0.027127 \rule[-3mm]{0mm}{7mm}\\ \hline
		\end{tabular}
\end{table}

\begin{figure}[htbp]
	\centering
	\includegraphics[width=7cm,height=5cm]{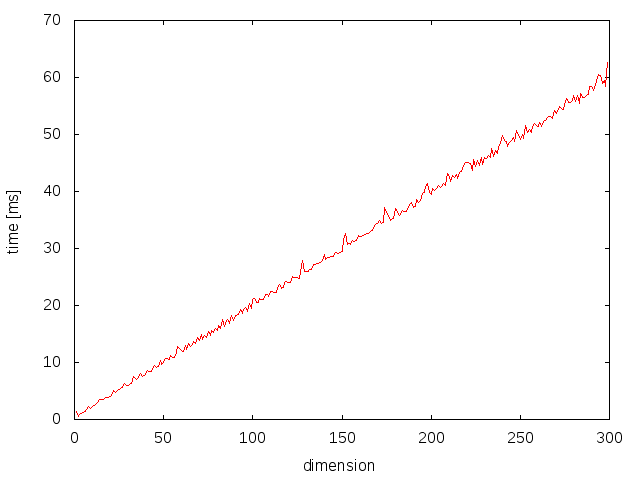}
	\caption{The computation time with respect to the dimensions}
	\label{time}
\end{figure}

The advantage of our method is its efficiency in high-dimensional cases. 
The calculation time of the normalizing constant $\mathcal{C}(\theta,\gamma)$ with multiple dimensions is demonstrated in Figure \ref{time}. 
The parameters $\theta$ and $\gamma$ are determined by the random number generator in C++ library and $N=200$. 

The calculation is very rapid and stable. 
The calculation time increases linearly. with the number of dimensions.
Since the derivatives of the normalizing constant are calculated with the same numerical integration formula, we prove that the derivatives can be calculated efficiently. 
The MLE algorithm, introduced in Section 4, is based on the gradient descent method. 
The bottleneck depends on the optimizing method because the calculation of the gradients is not expensive. 
Since the gradient descent method converges linearly, it is not efficient enough for high-dimensional data.
This problem can be solved using the quasi-Newton method.   

\subsection{MLE}
\label{MLE}
In this section, some numerical experiments on the MLE using data with multiple dimensions are shown. The data are obtained by rejection sampling, where $\Sigma$ is assumed to be diagonal, that is, $O$ is not estimated.

\subsubsection{2-dimensional data}
\label{2d}
\begin{figure}[htbp]
	\centering
	\includegraphics[width=7cm,height=5cm]{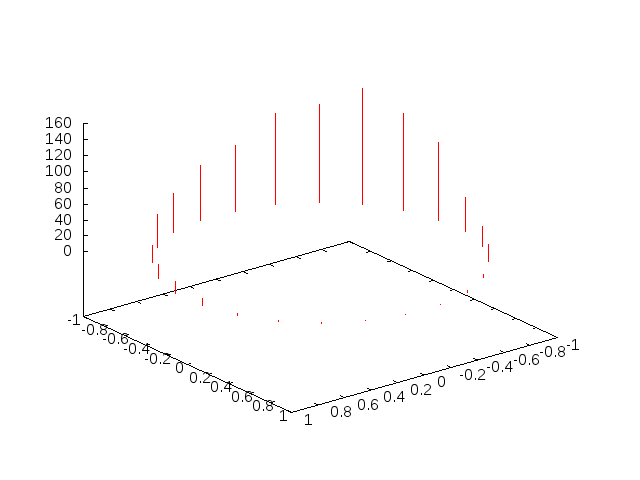}
	\caption{Histogram of the sampling points on a unit circle}
	\label{histgram}
\end{figure}

For data matrix $X=(x_1,x_2,\cdots,x_{1000}) \in \mathbb{R}^{2 \times 1000}$ (Figure \ref{histgram}), where
\begin{align*}
x_i &\overset{\text{iid}}{\sim} f(x;\theta^*,\gamma^*)=\frac{1}{\mathcal{C}(\theta^*,\gamma^*)}\exp \left(\sum_{i=1}^{2}(-\theta^*_i x_i^2+\gamma^*_i x_i)\right), \\
\theta^*&=  \left(
\begin{array}{c}
1.0   \\
2.0   
\end{array}
\right) \; \text{and} \; 
\gamma^*=  \left(
\begin{array}{c}
1.0   \\
2.0   
\end{array}
\right),\\
\end{align*}
the following MLE values
\begin{align*}
\hat{\theta}&=  \left(
\begin{array}{c}
1.04519  \\
1.95468  
\end{array}
\right) \; \text{and} \; 
\hat{\gamma}=  \left(
\begin{array}{c}
1.04643   \\
2.01098   
\end{array}
\right)\\
\end{align*}
are obtained.

\subsubsection{3-dimensional data}
\label{3d}
\begin{figure}[htbp]
	\centering
	\includegraphics[width=7cm,height=5cm]{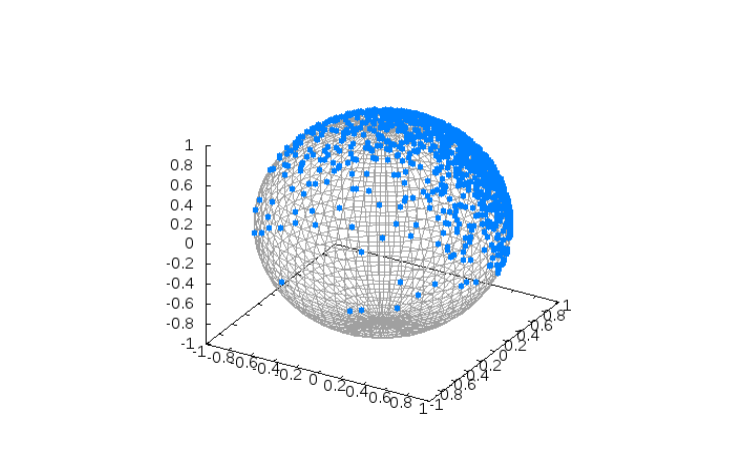}
	\caption{Sampling points on a unit sphere}
	\label{sample}
\end{figure}

For a data matrix $X=(x_1,x_2,\cdots,x_{1000}) \in \mathbb{R}^{3 \times 1000}$ (Figure \ref{sample}), where
\begin{align*}
x_i &\overset{\text{iid}}{\sim} f(x;\theta^*,\gamma^*)=\frac{1}{\mathcal{C}(\theta^*,\gamma^*)}\exp \left(\sum_{i=1}^{3}(-\theta^*_i x_i^2+\gamma^*_i x_i)\right), \\
\theta^*&=  \left(
\begin{array}{c}
1.0   \\
2.0   \\
3.0
\end{array}
\right) \; \text{and} \; 
\gamma^*=  \left(
\begin{array}{c}
1.0   \\
2.0   \\
3.0
\end{array}
\right),\\
\end{align*}
the following MLE values
\begin{align*}
\hat{\theta}&=  \left(
\begin{array}{c}
1.02215  \\
1.99636   \\
2.98587
\end{array}
\right) \; \text{and} \; 
\hat{\gamma}=  \left(
\begin{array}{c}
0.90462   \\
2.01793   \\
2.99908
\end{array}
\right)\\
\end{align*}
are obtained.

\subsubsection{10-dimensional data}
\label{10d}
For a data matrix $X=(x_1,x_2,\cdots,x_{1000}) \in \mathbb{R}^{10 \times 1000}$, where
\begin{align*}
x_i \overset{\text{iid}}{\sim} &f(x;\theta^*,\gamma^*)\\
=&\frac{1}{\mathcal{C}(\theta^*,\gamma^*)}\exp \left(\sum_{i=1}^{10}(-\theta^*_i x_i^2+\gamma^*_i x_i)\right), \\
\theta^*&=(1,2,3,4,5,6,7,8,9,10)^T ,\\
\gamma^*&=(1,2,3,4,5,6,7,8,9,10)^T
\end{align*}
the following MLE values
\begin{align*}
\hat{\theta}=(&1.2849,2.6223,3.0729,4.3598,5.0906,\\
&5.6851,6.2268,7.2384,7.9931,9.0433)^T ,\\
\hat{\gamma}=(&0.9962,2,1081,2,9225,4.1555,4.8213,\\
&5.6873,6.3327,7.4991,8.1197,9.2781)^T
\end{align*}
are obtained.

\section{Application of Fisher--Bingham distribution to S-VAE}
\label{application}
\subsection{Introduction to S-VAE}
\label{svae}
As mentioned in Section 5, our method enables to perform the MLE with high-dimensional data which cannot be achieved using other methods such as the holonomic gradient method.
Because of this advantage, we can apply our method to the MLE of the latent variables of variational auto-encoders (VAE). 

It is essential to comprehend the auto-encoders (AE) mechanism before introducing the VAE. 
AE is a generating model with two networks. 
One is called the encoder, while the other is called the decoder. 
The encoder transforms images into vectors with smaller dimensions than those of the images. 
For example, MNIST is a dataset with handwritten numbers from 0 to 9, and each image has 28 $\times$ 28 dimensions. 
With the encoder of the AE, an image with 28 $\times$ 28 dimensions is transformed into a vector with 2-20 dimensions. 
If an image becomes complicated, the dimensions of the latent space increase. 
For instance, the latent space of human face data has about 100 dimensions. 

\begin{figure*}[htbp]
	\centering
	\includegraphics[width=16cm,height=6cm]{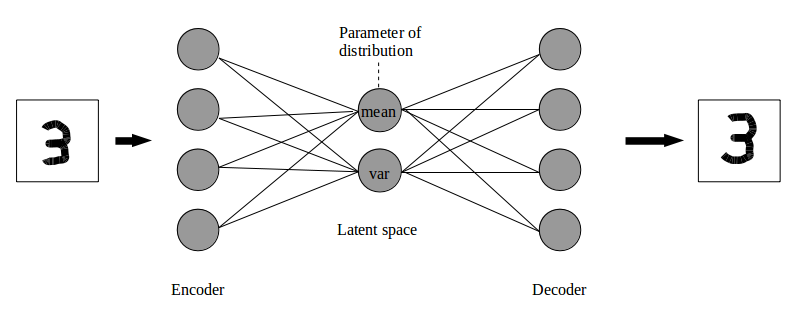}
	\caption{Auto-encoder}
	\label{ae}
\end{figure*}

VAE is a generating model derived from AE.
The difference between the VAE and AE is that the VAE assumes that the vectors obtained from images are generated from some distributions (\cite{doersch2016tutorial,kingma2013auto}). 
The encoder estimates the parameters of distributions, while the decoder generates an image from the vector, which is a sample of the estimated distributions.
For example, in most cases, the vectors are assumed to be generated from some normal distributions in Euclidean space.
In this section, the VAE implies that the normal distribution in Euclidean space is assumed.
Considering MNIST, 10 normal distributions match with different numbers from 0 to 9. 
If we decode a vector obtained from the distribution of zero, something like zero will probably be generated. 
Therefore, the encoder of the VAE transforms images into parameters, such as the means and variations of the distributions, instead of vectors themselves. 
The advantage of the VAE compared to the AE the VAE that it gives the structure of the latent variables. 

\begin{figure*}[htbp]
	\centering
	\includegraphics[width=16cm,height=6cm]{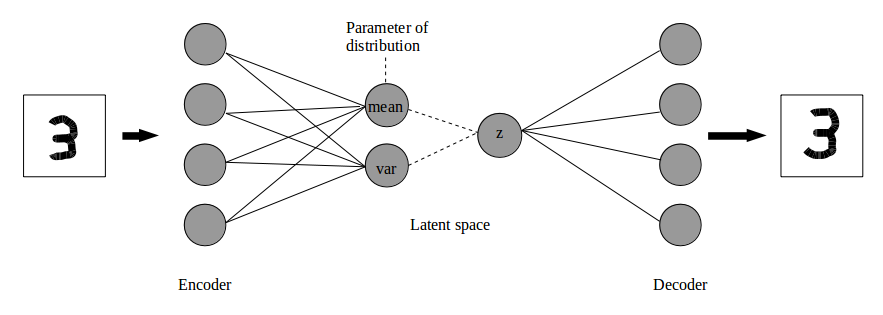}
	\caption{Variational auto-encoder}
	\label{vae}
\end{figure*}

S-VAE is a VAE assuming that the vectors are generated from distributions restricted to a high-dimensional unit hypersphere. Moreover, the vectors are assumed to be generated from the von Mises--Fisher distribution. 

\begin{definition}
	For a $p$-dimensional multivariate distribution, the von Mises--Fisher distribution is given by the density function
	\begin{align*}
	f(x;\mu,\kappa):=\frac{1}{\mathcal{C}(\kappa)}\exp{(\kappa \mu^T x)} \mathrm{d}_{\mathcal{S}^{p-1}}(x),
	\end{align*}
	where $x \in \mathbb{R}^p$ and
	\begin{align*}
	\mathcal{C}(\kappa):=\int_{\mathcal{S}^{p-1}}\exp{\left(\kappa \mu^T x\right)} \mathrm{d}_{\mathcal{S}^{p-1}}(x)
	\end{align*}
	is the normalizing constant and 
	$\mathrm{d}_{\mathcal{S}^{p-1}}(x)$ is the uniform measure in the $(p-1)$-dimensional sphere $\mathcal{S}^{p-1}$.
\end{definition}

The von Mises--Fisher distribution is a special case of the Fisher--Bingham distribution when restricting $\theta_i=0$ for all $i$. 
When $\kappa=0$, the von Mises--Fisher distribution becomes a uniform distribution on a unit sphere. 

\cite{davidson2018hyperspherical} proposed two reasons to assume that the latent space is a unit sphere rather than Euclidean space. 
First, S-VAE allows for a uniform distribution on the hypersphere to be a prior,  which is a truly uninformative prior. 
Since uniform distribution in Euclidean space does not exist, the VAE must make some informative assumptions about the prior. 
Second, with high-dimensional data, "the soap-bubble-effect", where the latent variables converge on a hyperspherical shell, is observed. 
Therefore, the assumption of the latent space as a Euclidean hyperplane would be improper because $\mathbb{R}^p$ is not homeomorphic to $\mathcal{S}^{p-1}$.

\begin{figure}[htbp]
	\centering
	\includegraphics[width=4cm,height=4cm]{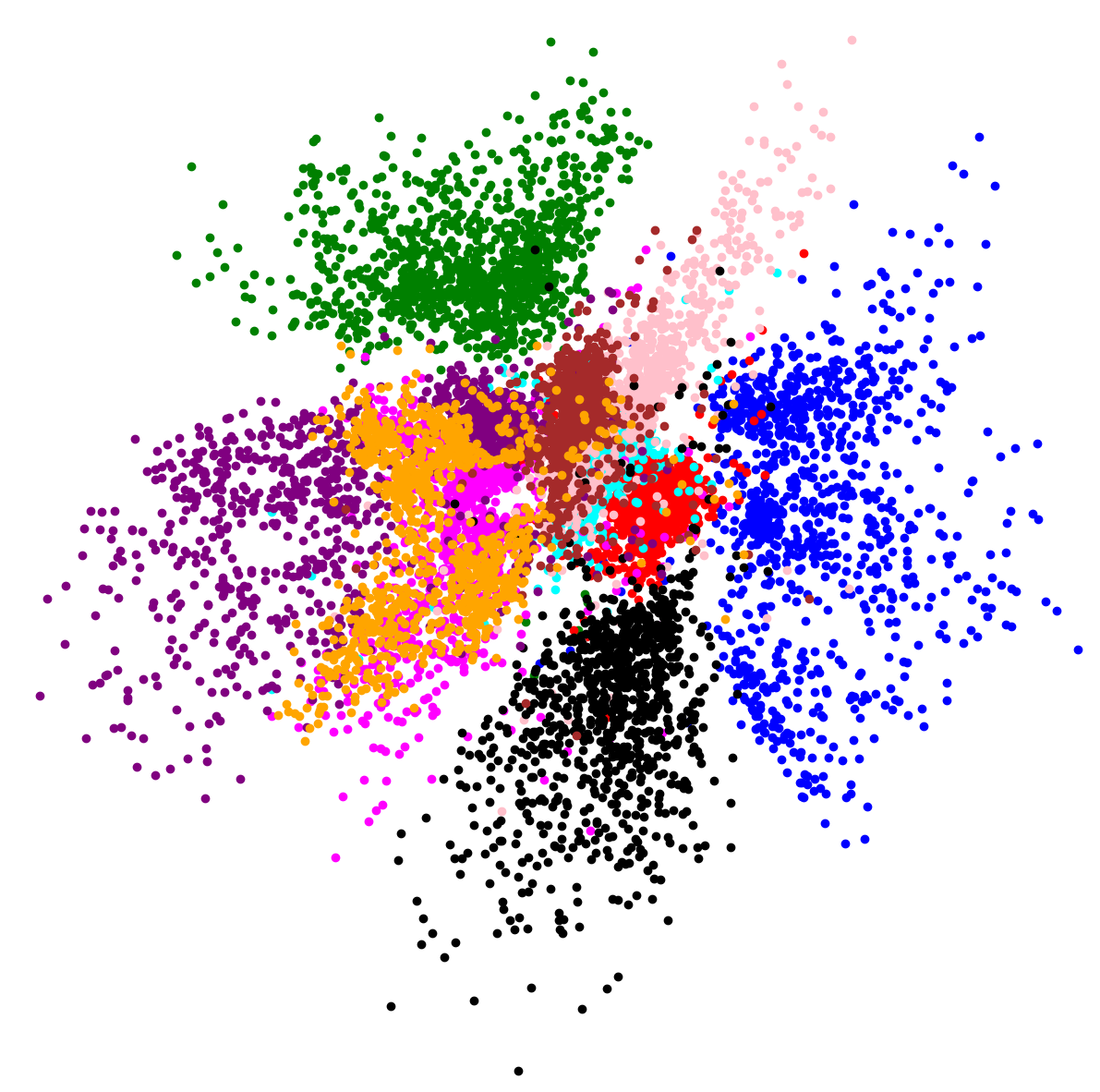}
	\caption{Latent space of the VAE (\cite{davidson2018hyperspherical})}
	\label{cluster of VAE}
\end{figure}

\begin{figure}[htbp]
	\centering
	\includegraphics[width=6cm,height=4cm]{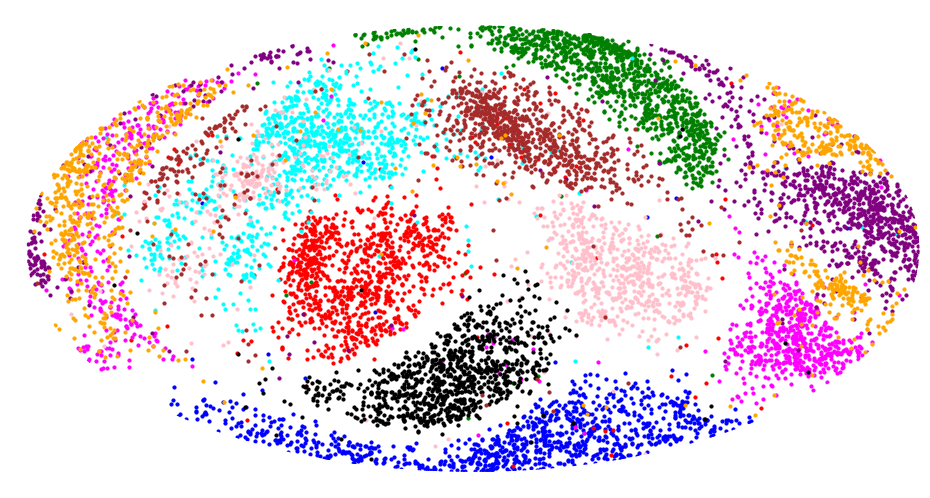}
	\caption{Latent space of the S-VAE (\cite{davidson2018hyperspherical})}	
	\label{cluster of SVAE}
\end{figure}

The AE, VAE and S-VAE are unsupervised machine learning, which means that the labels of images are not used for the training of networks.
However, images with high similarities, such as several handwritten images of the number one, become close to each other in latent space.
Therefore, some clusters are generated in latent space as shown in Figures \ref{cluster of VAE} and \ref{cluster of SVAE} (\cite{davidson2018hyperspherical}).
Each color matches with each label from 0 to 9. 
Figure \ref{cluster of SVAE} is the Hammer projection, a projection that shows a sphere on a 2-dimensional plain.
Therefore, the latent variable representation of each image can be obtained with a pre-trained model.
The parameters of these clusters can then be estimated using MLE.
The algorithm mentioned in Section 4 can be applied specifically to S-VAE. 

In addition, the conditional VAE (CVAE) is a supervised model that uses the label information to train the networks.

\begin{figure*}[htbp]
	\centering
	\includegraphics[width=16cm,height=7cm]{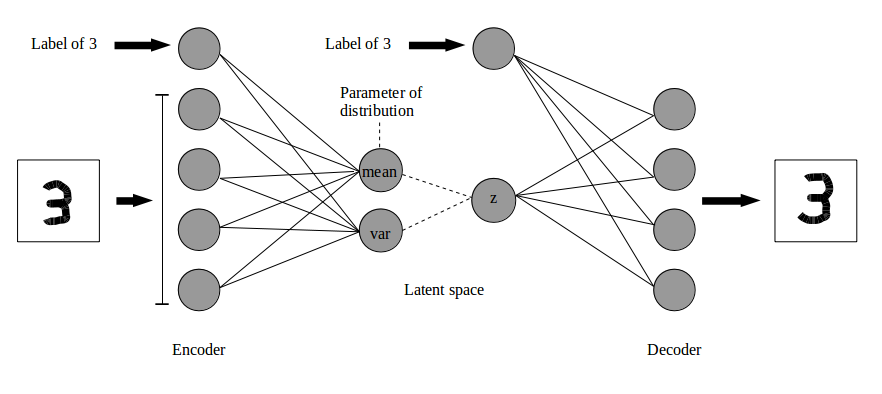}
	\caption{The conditional variational auto-encoder}
	\label{cvae}
\end{figure*}

It is common to consider adding more details to the labels of CVAE.
For example, if a pre-trained CVAE with human faces is labeled with continents, such as Asia, Europe and Africa, we want to add the details of continents to the labels.
Additionally, we must consider the situation in which a pre-trained CVAE with images of fish labeled with different species. 
When a new species is discovered, we may want to add this new species label to the network.
In most cases, it is common to train the networks with the dataset to which new labels are appended.
However, the computing cost is high. 
For example, the training of the VAE of MNIST with the latent space of a $\mathcal{S}^2$ using a CPU (Dell latitude 5480) takes several hours in our experiment.

If the latent variables are assumed to be generated with Fisher--Bingham distributions, the distribution of the images with new labels can be obtained by estimating its parameters via MLE.
In this way, since the computation cost of MLE is much lower than the training process of CVAE, new labels can be appended much more easily.
In the VAE of MNIST, the estimation of the parameters takes only seconds.
Therefore, the MLE algorithm mentioned in Section 4 can be applied for S-VAE.
Attention should be paid to other methods for the MLE of the Fisher--Bingham distribution, such as the holonomic gradient method and the saddlepoint approximation method, are not appropriate for high-dimensional data. 

\subsection{Application results}
\label{result}
In this section, the handwritten images dataset MNIST is used to train S-VAE. 
The experiments with the latent space $\mathcal{S}^2$ and $\mathcal{S}^5$ are shown. 
The models are trained according to \cite{davidson2018hyperspherical} \footnote{https://github.com/nicola-decao/s-vae-pytorch}.
It is necessary to add a sigmoid layer to the decoder. 
The training process is increased from 1000 to 2000 epochs with early-stopping at 50. 
Besides these changes, the training procedures are mostly the same as those mentioned by \cite{davidson2018hyperspherical}. 
Note that the experiment results are not accurate enough. 
In this section, the accuracy of the model means roughly the degree of the similarity of generated images to the handwritten numbers. 
The similarity is judged by the recognition of the authors. 
However, some concerns remain, such as the accuracy of the models and that of the Fisher--Bingham distribution assumption.

\subsubsection{Latent space $\mathcal{S}^2$}
\label{S2}
\begin{figure}[htbp]
	\centering
	\includegraphics[width=8cm,height=2cm]{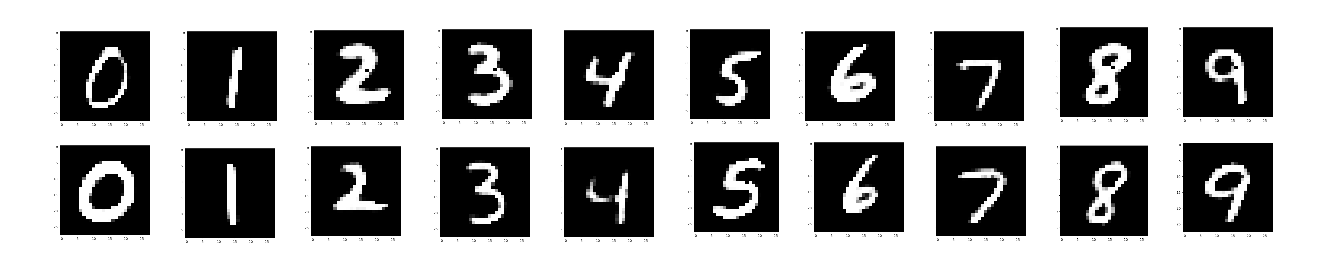}
	\caption{Images generated with the $\mathcal{S}^2$-VAE}
	\label{2g}
\end{figure}

The images generated by S-VAE are shown in Figure \ref{2g}.

The latent space is assumed to be a 2-dimensional unit sphere to visualize the latent space easily.
However, in terms of the accuracy of the model, it is probably not enough to learn the features of $28 \times 28$ dimensional images.
Therefore, the images generated by the model are not accurate; specifically, the features of the handwritten 2, 4, and 5 are not learned completely. 

\begin{figure}[htbp]
	\centering
	\includegraphics[width=8cm,height=6cm]{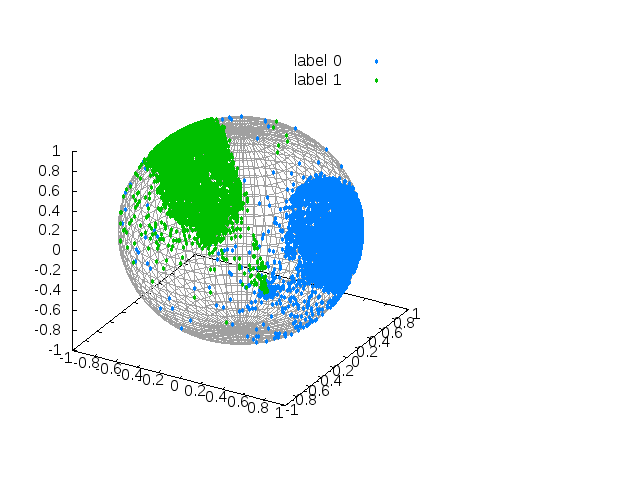}
	\caption{cluster of label 0 and 1}
	\label{distribution1}
\end{figure}

\begin{figure}[htbp]
	\centering
	\includegraphics[width=8cm,height=6cm]{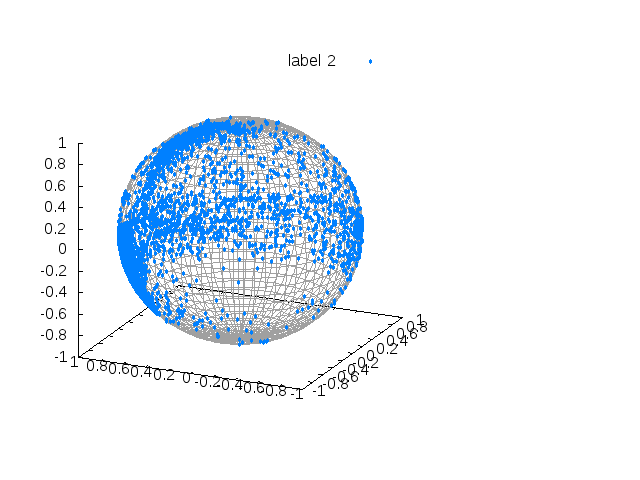}
	\caption{cluster of label 2}
	\label{distribution2}
\end{figure}

For example, the distribution of the latent variables of labels 0 and 1 is illustrated in Figure \ref{distribution1}.
It is clear that the areas of label 0 and label 1 are separate, and the data are accumulating.
However, the latent variables of label 2 are spread on the sphere, as shown in Figure \ref{distribution2}.

Given the latent variables mentioned above, the following MLE values for each label are obtained and listed in Table \ref{tab:MLE}.

\begin{table*}[htb]
	\caption{The MLE values of $\mathcal{S}^2$-VAE}
	\label{tab:MLE}
		\centering
	\begin{tabular}{ccccccccccc}
		label&0 & 1 & 2 & 3 &4 \rule[-2mm]{0mm}{6mm}  \\ \hline
		$\hat{\theta}$ &$ \left(
		\begin{array}{c}
		11.1187\\
		6.1277\\
		12.7461
		\end{array}
		\right)$& $ \left(
		\begin{array}{c}
		9.4918 \\
		7.1799 \\
		13.3121
		\end{array}
		\right)$ & $ \left(
		\begin{array}{c}
		10.6758 \\
		9.1414 \\
		10.1699
		\end{array}
		\right)$&$ \left(
		\begin{array}{c}
		11.5702 \\
		8.5393 \\
		9.8738
		\end{array}
		\right)$&$ \left(
		\begin{array}{c}
		9.8826 \\
		9.6800 \\
		10.4312
		\end{array}
		\right)$ \rule[-8mm]{0mm}{18mm}  \\  \hline
		$\hat{\gamma}$ & $ \left(
		\begin{array}{c}
		5.5963 \\
		2.1342 \\
		-3.6421
		\end{array}
		\right)$ & $ \left(
		\begin{array}{c}
		-2.2054 \\
		-1.7684 \\
		11.8523
		\end{array}
		\right)$ & $ \left(
		\begin{array}{c}
		-1.6375\\
		0.8448\\
		0.3313
		\end{array}
		\right)$ & $ \left(
		\begin{array}{c}
		4.1036\\
		0.6904\\
		3.1363
		\end{array}
		\right)$ & $ \left(
		\begin{array}{c}
		-1.6362 \\
		-1.9930 \\
		-1.3880
		\end{array}
		\right)$ \rule[-8mm]{0mm}{18mm} \\ \hline
		label&5 & 6 & 7 & 8 &9 \rule[-2mm]{0mm}{7mm} \\  \hline
		$\hat{\theta}$ &$ \left(
		\begin{array}{c}
		8.5678\\
		9.6602\\
		11.7626
		\end{array}
		\right)$& $ \left(
		\begin{array}{c}
		12.1134\\
		10.3569\\
		7.5188
		\end{array}
		\right)$ & $ \left(
		\begin{array}{c}
		9.7964\\
		10.9692\\
		9.2246
		\end{array}
		\right)$&$ \left(
		\begin{array}{c}
		10.0447\\
		8.4945\\
		11.4392
		\end{array}
		\right)$&$ \left(
		\begin{array}{c}
		9.4076\\
		10.1294\\
		10.4539
		\end{array}
		\right)$ \rule[-8mm]{0mm}{18mm} \\ \hline
		$\hat{\gamma}$ & $ \left(
		\begin{array}{c}
		2.9471\\
		-0.8795\\
		0.1349
		\end{array}
		\right)$ & $ \left(
		\begin{array}{c}
		5.1722\\
		0.1787\\
		-2.8050
		\end{array}
		\right)$ & $ \left(
		\begin{array}{c}
		-2.4272\\
		-1.5758\\
		-1.3331
		\end{array}
		\right)$ & $ \left(
		\begin{array}{c}
		-0.3664 \\
		0.9943 \\
		3.0225
		\end{array}
		\right)$ & $ \left(
		\begin{array}{c}
		-1.7629 \\
		-2.5092 \\
		-1.9080
		\end{array}
		\right)$ \rule[0mm]{0mm}{10mm}
	\end{tabular}
\end{table*}

\begin{figure}[htbp]
	\centering
	\includegraphics[width=8cm,height=2cm]{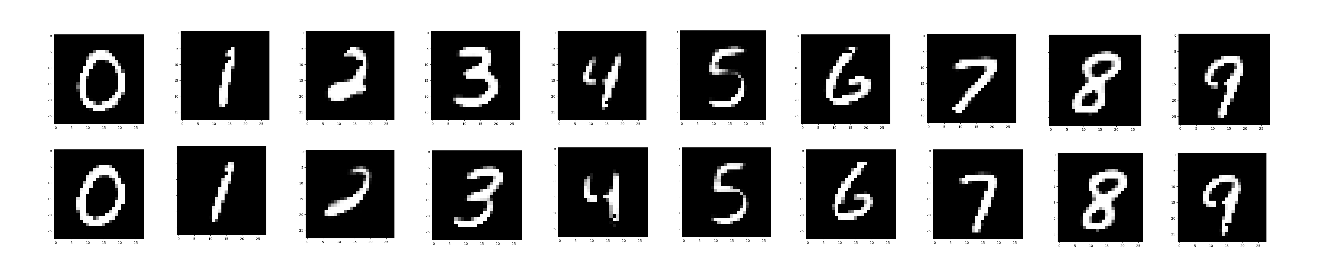}
	\caption{Images generated with the MLE on $\mathcal{S}^2$}
	\label{2gs}
\end{figure}

The images generated with the samples of the distributions with the above parameters are shown in Figure \ref{2gs}

The estimation works better with data accumulating as a cluster. 
For example, the images generated with the MLE of label 1 appear highly similar to the handwritten images of 1. 
However, if the data is spread on the sphere, such as label 2, the images
generated from the estimations will either appear like other numbers or not like a number at all.
In our opinion, this problem occurs because the training procedure is insufficient. 
The estimation should perform more accurately with a better model that distributes different labels in different clusters.   
The dimension of $\mathcal{S}^2$ may not be suitable for the training process.  

\subsubsection{Latent space $\mathcal{S}^5$}
\label{S5}
\begin{figure}[htbp]
	\centering
	\includegraphics[width=8cm,height=2cm]{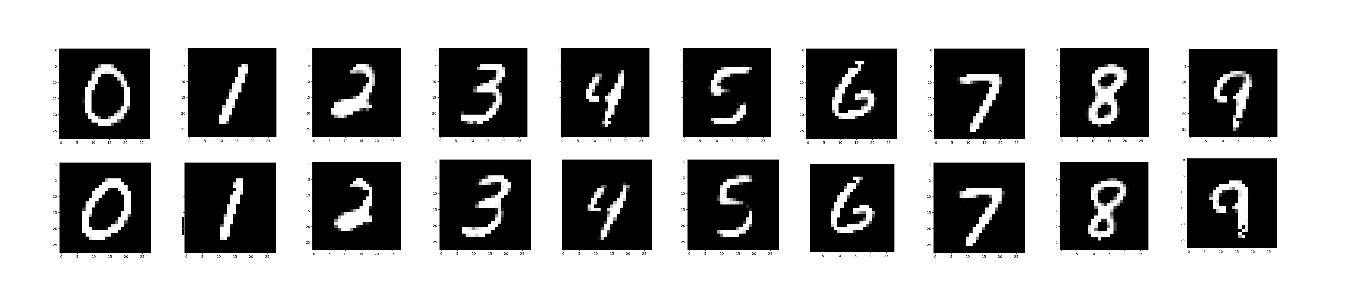}
	\caption{Images generated with the $\mathcal{S}^5$-VAE}
	\label{5g}
\end{figure}

The images generated by S-VAE are shown in Figure \ref{5g}

The latent space is assumed to be a 5-dimensional unit sphere, and thus, cannot be visualized. 
Nevertheless, the accuracy of the model is improved such that the generated images are closer to the handwritten numbers. 
The features of each number are learned more completely than assuming the latent space as $\mathcal{S}^2$.

With latent variables calculated by the encoder, the following MLE values for each label are obtained and listed in Table \ref{tab:MLE5}.

\begin{table*}[htb]
	\caption{The MLE values of $\mathcal{S}^5$-VAE}
		\label{tab:MLE5}
	\centering
	\begin{tabular}{ccccccccccc}
		label&0 & 1 & 2 & 3 &4 \rule[-2mm]{0mm}{6mm}  \\ \hline
		$\hat{\theta}$ &$ \left(
		\begin{array}{c}
		7.0028\\
		9.4078\\
		11.8385\\
		11.4298\\
		11.5790\\
		8.7433
		\end{array}
		\right)$& $ \left(
		\begin{array}{c}
		8.7665\\
		21.1636\\
		12.6326\\
		4.9772\\
		6.0491\\
		6.4194
		\end{array}
		\right)$ & $ \left(
		\begin{array}{c}
		8.6867\\
		13.1393\\
		10.3493\\
		8.5949\\
		9.6434\\
		9.5842
		\end{array}
		\right)$&$ \left(
		\begin{array}{c}
		10.2006\\
		12.6433\\
		9.1409\\
		9.7167\\
		8.4137\\
		9.8842
		\end{array}
		\right)$&$ \left(
		\begin{array}{c}
		7.9105\\
		9.1753\\
		7.9465\\
		13.1474\\
		10.1395\\
		11.6815
		\end{array}
		\right)$ \rule[-14mm]{0mm}{30mm}  \\  \hline
		$\hat{\gamma}$ & $ \left(
		\begin{array}{c}
		-3.7674\\
		-4.6981\\
		-0.9165\\
		-1.4294\\
		-2.7827\\
		0.6998
		\end{array}
		\right)$ & $ \left(
		\begin{array}{c}
		4.2662\\
		12.6101\\
		-3.7816\\
		1.6432\\
		-2.8957\\
		0.7323
		\end{array}
		\right)$ & $ \left(
		\begin{array}{c}
		-1.7124\\
		4.2293\\
		3.3046\\
		2.4545\\
		0.6192\\
		0.2538
		\end{array}
		\right)$ & $ \left(
		\begin{array}{c}
		1.5651\\
		-5.4851\\
		2.4474\\
		3.7723\\
		-0.0504\\
		1.2384
		\end{array}
		\right)$ & $ \left(
		\begin{array}{c}
		0.8373\\
		2.3149\\
		-2.5721\\
		-3.4944\\
		0.6412\\
		-5.8691
		\end{array}
		\right)$ \rule[-14mm]{0mm}{30mm}  \\ \hline
		label&5 & 6 & 7 & 8 &9 \rule[-2mm]{0mm}{8mm} \\  \hline
		$\hat{\theta}$ &$ \left(
		\begin{array}{c}
		11.6096\\
		8.7744\\
		10.3410\\
		10.6990\\
		9.8872\\
		8.6861
		\end{array}
		\right)$& $ \left(
		\begin{array}{c}
		8.9869\\
		9.7802\\
		10.1568\\
		12.3415\\
		13.0652\\
		5.6728
		\end{array}
		\right)$ & $ \left(
		\begin{array}{c}
		9.7467\\
		8.4359\\
		10.3852\\
		9.1947\\
		10.2960\\
		11.9392
		\end{array}
		\right)$&$ \left(
		\begin{array}{c}
		10.0842\\
		11.2982\\
		9.6019\\
		11.4131\\
		7.2654\\
		10.3349
		\end{array}
		\right)$&$ \left(
		\begin{array}{c}
		9.2238\\
		8.7748\\
		8.3899\\
		13.0000\\
		7.6813\\
		12.9272
		\end{array}
		\right)$ \rule[-14mm]{0mm}{30mm} \\ \hline
		$\hat{\gamma}$ & $ \left(
		\begin{array}{c}
		1.9250\\
		-2.6234\\
		2.5431\\
		-0.4646\\
		-0.2525\\
		-0.1028
		\end{array}
		\right)$ & $ \left(
		\begin{array}{c}
		-4.3087\\
		2.6298\\
		2.6724\\
		-5.0590\\
		-5.9750\\
		1.6123
		\end{array}
		\right)$ & $ \left(
		\begin{array}{c}
		1.5073\\
		1.1309\\
		-1.0298\\
		-3.2854\\
		1.5368\\
		1.5781
		\end{array}
		\right)$ & $ \left(
		\begin{array}{c}
		2.0016\\
		-1.6481\\
		0.3854\\
		3.5668\\
		1.5604\\
		-2.0743
		\end{array}
		\right)$ & $ \left(
		\begin{array}{c}
		1.8123\\
		-0.0305\\
		-1.7664\\
		-3.3737\\
		1.3602\\
		-3.0582
		\end{array}
		\right)$ \rule[-14mm]{0mm}{30mm} 
	\end{tabular}
\end{table*}

\begin{figure}[htbp]
	\centering
	\includegraphics[width=8cm,height=2cm]{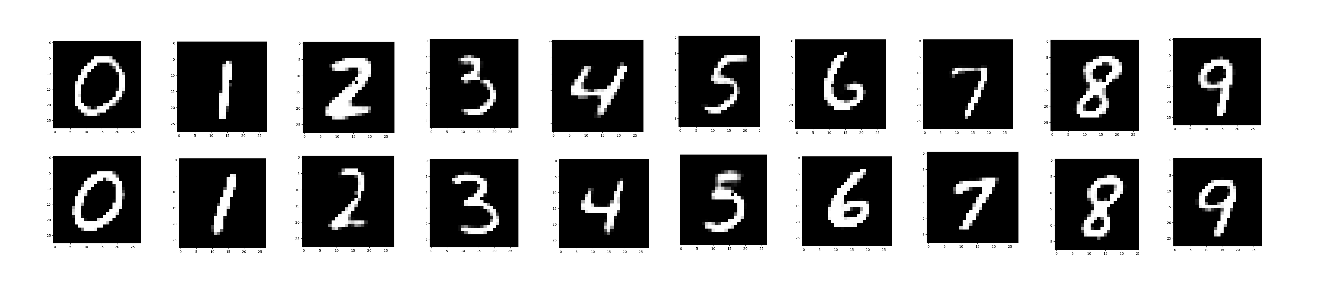}
	\caption{Images generated with the MLE on $\mathcal{S}^5$}
	\label{5gs}
\end{figure}

The images generated with the samples of the distributions with the above parameters are shown in Figure \ref{5gs}.

\section{Concluding remarks}
\label{concluding}
In this paper, the continuous Euler transform was applied to the calculation of the normalizing constant of the Fisher--Bingham distribution and its derivatives.
The Fourier transform can represent the normalizing constant and the derivatives.
In this way, the Fourier transform representation can be efficiently calculated with the continuous Euler method. 
This approach can achieve any accuracy with a low computation cost. 

With this method, the MLE can be performed rapidly, even with high-dimensional data. 
The numerical experiments with data of different dimensions generated by rejection sampling are shown. 
With the gradient descent method, the data can be calculated in 10 dimensions within a few minutes. 
If data with higher dimensions are given, it is necessary to change the algorithm to the quasi-Newton method so that the computation can be more efficient. 
The second-order partial derivatives can also be obtained using the continuous Euler method.

The MLE of the Fisher--Bingham distribution can be applied to the estimation of the latent variables of S-VAE.
S-VAE is a generating model restricting the latent space as a unit sphere, where the latent variables are assumed to be generated from distributions on the sphere. 
When considering the latent variables of S-VAE, if the images are simple, such as handwritten numbers in the MNIST dataset, the latent space can be taken as 2 to 10 dimensions.
Therefore, the proposed algorithm can be used. 
If the images are complicated, such as human faces and animals, the algorithm with the quasi-Newton method is required.


%
%

\begin{acknowledgements}
We are grateful to Taichi Kiwaki for providing access to a GPU and giving advice about VAE. 
We thank Kazuki Matoya for general discussion about the application of MLE to VAE. 
We would like to thank Shun Sato for providing some advice about numerical computation. 
This work was supported by all members in the mathematical informatics 3rd laboratory of the University of Tokyo. 
Finally, we appreciate Tomonari Sei, who joined our discussion and gave a lot of advice. 
Ken'ichiro Tanaka is supported by the grant-in-aid of Japan Society of the Promotion of Science with KAKENHI Grant Number 17K14241. 
\end{acknowledgements}

%
%

\bibliographystyle{spbasic}      
\bibliography{template.bib}   

%
%

\end{document}